\documentclass[10pt,floatfix]{revtex4}
\usepackage{bm}
\usepackage{amsmath}
\usepackage{graphics}
\def\edth{\;\raise1.0pt\hbox{$'$}\hskip-6pt\partial}
\def\baredth{\;\overline{\raise1.0pt\hbox{$'$}\hskip-6pt
\partial}}
\newcommand{\vc}[1]{\ensuremath{\bm{#1}}}
\newcommand{\vh}[1]{\ensuremath{\hat{\bm{#1}}}}
\newcommand{\avg}[1]{\ensuremath{\langle #1 \rangle}}
\newcommand{\ThreeJ}[6]{\left (\begin{array}{ccc}#1&#2&#3\\
#4&#5&#6\end{array} \right )}

\newcommand{\even}{\ensuremath{\epsilon}}
\newcommand{\odd}{\ensuremath{\beta}}
\newcommand{\fwhm}{\ensuremath{\theta_{\text{FWHM}}}}
\newcommand{\myfigure}[1]{\resizebox{3in}{!}{\includegraphics{#1}}}
\newcommand{\Pol}{\bm{\mathcal P}}
\newcommand{\vp}{\bar{\vc{m}}}
\newcommand{\vm}{\vc{m}}
\newcommand{\myref}[5]{#1, #2~\textbf{#3}, #4 (#5).}
\newcommand{\AandA}{Astron. \& Astrophys.}

\begin{document}
\title{CMB Lensing Reconstruction on the Full Sky}
\date{\today}
\author{Takemi Okamoto}
\affiliation{Department of Physics and the Center for Cosmological Physics, 
University of Chicago, Chicago, IL 60637}
\email{tokamoto@oddjob.uchicago.edu}
\author{Wayne Hu}
\affiliation{Center for Cosmological Physics and the
Department of Astronomy and Astrophysics and the Enrico Fermi Institute, 
University of Chicago, Chicago, IL 60637}
\email{whu@background.uchicago.edu}
\begin{abstract}
Gravitational lensing of the microwave background by the intervening
dark matter mainly arises from large-angle fluctuations in the
projected gravitational potential and hence offers a unique
opportunity to study the physics of the dark sector at large scales.
Studies with surveys that cover greater than a percent of the sky will
require techniques that incorporate the curvature of the sky.  We lay
the groundwork for these studies by deriving the full sky minimum
variance quadratic estimators of the lensing potential from the CMB
temperature and polarization fields.  We also present a general
technique for constructing these estimators, with harmonic space
convolutions replaced by real space products, that is appropriate for
both the full sky limit and the flat sky approximation.  This also
extends previous treatments to include estimators involving the
temperature-polarization cross-correlation and should be useful for
next generation experiments in which most of the additional
information from polarization comes from this channel due to
sensitivity limitations.
\end{abstract}

\maketitle
\section{Introduction}
\label{Sect:Introduction}

Weak gravitational lensing of the microwave background anisotropies
offers a unique opportunity to study the dark matter and energy
distribution at intermediate redshifts and large scales.  In addition to
producing modifications in the CMB temperature and polarization power
spectra \cite{Seljak96,ZaldarriagaSeljak98}, lensing of the CMB fields
produces higher-order correlations between the multipole moments
\cite{Bernardeau97,Bernardeau98}.  Quadratic combinations of the CMB
fields can be used to form estimators of the projected gravitational
potential, and therefore of the projected mass
\cite{ZaldarriagaSeljak99, GuzikEtAl00}.  The minimum variance
quadratic estimator can in principle map the projected mass on large
angular scales out to multipole moments of $L\sim 10^2$
\cite{Hu01a,Hu01b} and contains nearly all of the information in the
higher moments of the lensed temperature field \cite{HirataSeljak02}.
Substantially more information lies in the lensed polarization fields
allowing high signal-to-noise lensing reconstruction and extending the
angular resolution out to $L\sim 10^3$ \cite{HuOkamoto02}.

Lensing reconstruction techniques involving the polarization fields
have previously only been developed for small surveys where the sky
can be taken to be approximately flat.  Since lensing is intrinsically
most sensitive to the projected potential at $L < 10^2$ or several
degrees on the sky, a treatment incorporating the curvature of the sky
is desirable.  In fact it is necessary for its application in removing
the lensing contaminant to gravitational wave polarization
\cite{HuOkamoto02,KnoxSong02,KesdenEtAl02} across large regions of the
sky.

We present a concise treatment of the effect of gravitational lensing
on CMB temperature and polarization harmonics in
Sect.~\ref{Sect:CMBLensingMultipoles}.  We construct the full sky
quadratic estimators of the lensing potential and compare their noise
properties to that for the flat sky expressions in
Sect.~\ref{Sect:QuadraticEstimators}.  We provide an efficient
algorithm for the construction of all estimators in
Sect.~\ref{Sect:AngularSpaceEstimators}.  
We summarize some useful properties of spin-weighted functions in
Appendix~\ref{Appendix:SpinSFunctions}.  Finally, we derive the flat sky
limits of the estimators and draw the connection to results in
\cite{HuOkamoto02} in Sect.~\ref{Appendix:FlatSky}.

\section{CMB Lensing in Multipole Space}
\label{Sect:CMBLensingMultipoles}

In this section, we give a pedagogical but concise derivation of the
lensing effect on the CMB temperature and polarization fields on the
sphere \cite{Hu00,ChallinorChon02}.  We emphasize the connections
between the formalism using spin-weighted spherical harmonics
\cite{ZaldarriagaSeljak97} and a tensorial approach
\cite{KamionkowskiEtAl97} which will be useful for the lensing
reconstruction in the following sections.

The temperature perturbation is characterized by a scalar function 
$\Theta(\vh{n})\equiv \Delta T(\vh{n})/T$, whose harmonic transform is 
given by
\begin{equation}
\Theta(\vh{n})\equiv \sum_{lm}\Theta_l^m Y_l^m(\vh{n}).
\label{Eqn:ThetaDecomposition}
\end{equation}
The polarization anisotropy of the microwave background is characterized by
a traceless, symmetric rank 2 tensor, which can be represented as 
(e.g. \cite{KamionkowskiEtAl97})
\begin{equation}
\Pol_{ij} = {}_{+2}A(\vh{n})\vp_i\vp_j + {}_{-2}A(\vh{n})\vm_i \vm_j,
\label{Eqn:PolTensor}
\end{equation}
where we have defined the complex Stokes parameters ${}_{\pm 2}A$ according to
\begin{eqnarray}
{}_{\pm 2}A(\vh{n}) & = & Q(\vh{n})\pm iU(\vh{n}).\label{Eqn:ComplexStokes}
\label{Eqn:ProjectionVector}
\end{eqnarray}
The spin projection vectors are given with respect to the measurement basis
$(\vh{e}_1,\vh{e}_2)$ by
\begin{eqnarray}
\vm &=& \frac{1}{\sqrt{2}}\left [\vh{e}_1 + i \vh{e}_2\right ],
\label{Eqn:vm}\\
\vp &=&\frac{1}{\sqrt{2}}\left [\vh{e}_1 - i \vh{e}_2\right ],
\label{Eqn:vp}
\end{eqnarray}
and form an eigenbasis under local rotations of basis vectors 
(see Appendix~\ref{Appendix:SpinSFunctions}).  In spherical polar coordinates,
$\vh{e}_1=\vh{e}_\theta$ and $\vh{e}_2=\vh{e}_\varphi$.
Under a local, right-handed rotation of the basis $(\vh{e}_1,\vh{e}_2)$ by an 
angle $\psi$, the complex Stokes parameters ${}_{\pm 2}A(\vh{n})$ acquire a 
phase $e^{\mp 2i\psi}$.  They act as spin-2 functions, with a corresponding
harmonic transform in terms of spin-weighted spherical harmonics \cite{Goldberg67}
given by \cite{ZaldarriagaSeljak97}
\begin{equation}
{}_{\pm 2}A(\vh{n})\equiv\sum_{lm}{}_{\pm 2}A_l^m {}_{\pm 2}Y_l^m(\vh{n}).
\label{Eqn:StokesDecomposition}
\end{equation}

A lens with a projected potential $\phi(\vh{n})$ maps the temperature
and polarization anisotropies according to 
\cite{BlanchardSchneider87,Bernardeau97,ZaldarriagaSeljak98}
\begin{eqnarray}
\Theta(\vh{n})&=&\tilde{\Theta}(\vh{n}+\nabla\phi(\vh{n}))\nonumber \\
&=&\tilde{\Theta}(\vh{n})+\nabla_i\phi(\vh{n})\nabla^i\tilde{\Theta}(\vh{n})
+{\mathcal O}(\phi^2),
\label{Eqn:TLensMap}\\
\Pol_{ij}(\vh{n})&=&\tilde{\Pol}_{ij}(\vh{n}+\nabla\phi(\vh{n}))\nonumber \\
&=&\tilde{\Pol_{ij}}(\vh{n})+\nabla_k\phi(\vh{n})\nabla^k
\tilde{\Pol_{ij}}(\vh{n})+{\mathcal O}(\phi^2),\label{Eqn:PLensMap}
\end{eqnarray}
where tildes denote the unlensed fields.  In the case of a weak gravitational
field under consideration,
lensing potential $\phi$ is obtained by a line-of-sight projection of the 
gravitational potential,
\begin{equation}
\phi(\vh{n}) = -2\int 
 d\eta \frac{\chi(\eta-\eta_s)}{\chi(\eta_s)\chi(\eta)}\Psi(\chi\vh{n},\eta),
\label{Eqn:LensingPotential}
\end{equation}
where $\eta$ is the conformal time, $\eta_s$ is the epoch of last scattering
and $\chi$ is the angular diameter distance in comoving coordinates.

Taking the harmonic transform of Eqn.~(\ref{Eqn:TLensMap}),
one readily shows that the 
the change to the temperature moments $\delta \Theta_l^m \equiv \Theta_l^m - \tilde \Theta_l^m$ 
are given by \cite{GolSpe99}
\begin{eqnarray}
\delta \Theta_l^m&\approx & \sum_{LM}\sum_{l'm'}\phi_L^M
\tilde{\Theta}_{l'}^{m'}I_{lLl'}^{mMm'},
\label{Eqn:LensedTlm}
\end{eqnarray}
with $I_{lLl'}^{mMm'}$ denoting the integral
\begin{eqnarray}
I_{lLl'}^{mMm'}&=&\int d\vh{n}Y_l^m{}^*
\nabla_iY_L^M\nabla^iY_{l'}^{m'}.
\label{Eqn:Iintegral}
\end{eqnarray}
The integral can be performed analytically using the relation
\cite{Varshalovich}
\begin{equation}
\int d\vh{n} {}_{s_1}Y_{l_1}^{m_1}(\vh{n}){}_{s_2}Y_{l_2}^{m_2}(\vh{n})
{}_{s_3}Y_{l_3}^{m_3}(\vh{n}) =
\frac{(-1)^{m_1+s_1}}{\sqrt{4\pi}}\left [\prod_{i=1}^3 2l_i+1\right ]^{1/2}
\ThreeJ{l_1}{l_2}{l_3}{-s_1}{-s_2}{-s_3}
\ThreeJ{l_1}{l_2}{l_3}{m_1}{m_2}{m_3}
\label{Eqn:ThreeJtosYlm}
\end{equation}
to yield
\begin{equation}
I_{lLl'}^{mMm'}=(-1)^m\ThreeJ{l}{L}{l'}{-m}{M}{m'}\, {}_0 F_{lLl'},
\label{Eqn:IResult}
\end{equation}
with the definition
\begin{equation}
{}_{\pm s}F_{lLl'}=\left [L(L+1)+l'(l'+1)-l(l+1)\right ]
\sqrt{\frac{(2L+1)(2l+1)(2l'+1)}{16\pi}}\ThreeJ{l}{L}{l'}{\pm s}{0}{\mp s}.
\label{Eqn:2FDefinition}
\end{equation}



The multipole expansion for the polarization fields 
proceeds by noting that ${}_{\pm 2}A(\vh{n})$ are the spin $\pm 2$ components
of the polarization tensor $\Pol_{ij}$.  Since the contraction with the spin 
projection vectors $\vm^i\vm^j$ projects out the spin 2 piece of a symmetric
tensor, the change in the complex Stokes parameters is given by
\begin{eqnarray}
\delta\left [{}_{+2}A(\vh{n})\right ]
&=& \vm^i\vm^j \delta\Pol_{ij}\nonumber \\
&\approx& \vm^i\vm^j\left [\nabla_k \tilde \Pol_{ij}(\vh{n})\right ]
\left [\nabla^k\phi(\vh{n})\right ].
\label{Eqn:LensingContribution}
\end{eqnarray}
The expression for the contribution to ${}_{-2}A(\vh{n})$ is obtained by 
replacing $\vm^i$ by $\vp^i$ in the above.
We denote the product $\vm^i\vm^j\nabla_k\tilde \Pol_{ij}$ using a spin-gradient
derivative $D_i$ (see Eq.~\ref{Eqn:GradientOp}), and write the lensing 
contribution as 
\begin{equation}
\delta\left [{}_{\pm 2}A(\vh{n})\right ] \approx
D^i\phi(\vh{n})
D_i[{}_{\pm 2}\tilde{A}(\vh{n})].
\label{Eqn:LensingContrib2}
\end{equation}
This relationship was given in \cite{Hu00} with the shorthand 
convention $D_i \rightarrow \nabla_i$ corresponding to the action of covariant
derivatives on the spin components of symmetric trace free tensors 
given in Eqn.~(\ref{Eqn:CovDerivFinal}) \cite{Goldberg67,ChallinorChon02}.
Expanding $\phi$ 
and ${}_{\pm 2}A$ in spin-weighted spherical harmonics and 
evaluating the inner product of their gradients using Eqn.~(\ref{Eqn:gradientdot}),
we obtain the lensing corrections
\begin{eqnarray}
\delta [{}_{\pm 2}A_l^m]& \approx & \sum_{LM}\sum_{l'm'}
\phi_L^M{}_{\pm 2}\tilde{A}_l^m{}_{\pm 2}I_{lLl'}^{mMm'},
\label{Eqn:LensedpmAlm}
\end{eqnarray}
where we define 
\begin{eqnarray}
{}_{\pm 2}I_{lLl'}^{mMm'}&=&(-1)^m\ThreeJ{l}{L}{l'}{-m}{M}{m'}
{}_{\pm 2}F_{lLl'}.\label{Eqn:pm2IDefinition}
\end{eqnarray}

We will be interested in the lensing expressions for the rotationally 
invariant combinations
\begin{eqnarray}
E_l^m &=&\frac{1}{2}\left [{}_{+2}A_l^m + {}_{-2}A_l^m\right ],
\label{Eqn:EMultipole} \\
B_l^m &=&\frac{1}{2i}\left [{}_{+2}A_l^m - {}_{-2}A_l^m\right ]
\label{Eqn:BMultipole},
\end{eqnarray}
which are the curl-free (``E-mode'') and gradient-free (``B-mode'') components 
of the polarization field.
>From the expressions (\ref{Eqn:LensedTlm}) and (\ref{Eqn:LensedpmAlm}), we find
the general expression for a lensed multipole moment to be
\begin{eqnarray}
\delta X_l^m&\approx& \sum_{LM}\sum_{l'm'}\phi_L^M(-1)^m
\ThreeJ{l}{l'}{L}{m}{-m'}{-M}{}_{s_X}F_{lLl'}
\left [\even_{ll'L}X_{l'}^{m'}+\odd_{ll'L}\bar X_{l'}^{m'} \right ],
\label{Eqn:LensedXlm}
\end{eqnarray}
where $X_l^m$ may be multipole moments of $\Theta$, $E$, or $B$, and
\begin{eqnarray}
\even_{ll'L}&=&\frac{1+(-1)^{L+l+l'}}{2},\nonumber \\
\odd_{ll'lL}&=&\frac{1-(-1)^{L+l+l'}}{2i}
\label{Eqn:EvenOddDefns}
\end{eqnarray}
ensure that the associated terms are nonzero only when $L+l+l'$ is even or odd,
respectively.  $\bar X$ denotes the parity complement of $X$, i.e. 
$\bar\Theta=0$, $\bar E = -B$, $\bar B = E$.
%

\section{Quadratic Estimators}
\label{Sect:QuadraticEstimators}
Lensing of the CMB fields mixes different multipoles through the convolution
(\ref{Eqn:LensedXlm}), and therefore correlates modes across a band determined by
the power in the deflection angles \cite{Hu00}.
The unlensed CMB multipoles $\tilde{X}_l^m$ are assumed to be Gaussian and 
statistically isotropic, so that the statistical properties are characterized
by diagonal covariances or power spectra
\begin{equation}
\avg{\tilde X_l^{m*}\tilde X'{}_{l'}^{m'}} = \delta_{ll'}\delta_{mm'}\tilde{C}_l^{XX'}.
\label{Eqn:UnlensedCl}
\end{equation}
The assumption of parity invariance implies that 
$\tilde{C}_l^{\Theta B}=\tilde{C}_l^{EB}=0.$  
The lensing potential is also assumed to be statistically isotropic so that
\begin{equation}
l(l+1) \avg{\phi_l^{m*} \phi_{l'}^{m'}} = \delta_{ll'}\delta_{mm'}{C}_l^{dd}.
\label{Eqn:phiCl}
\end{equation}
where we have multiplied through by $l(l+1)$ to reflect the weighting of
deflection angles.

It follows then that the lensed multipoles are also 
statistically isotropic with power spectra
\begin{equation}
\avg{X_l^{m*}X'{}_{l'}^{m'}} = \delta_{ll'}\delta_{mm'}C_l^{XX'}.
\label{Eqn:TotalCl}
\end{equation}
Since ${X}_l^m$ denotes the measured multipoles, the power spectra 
contain all sources to the variance, including detector noise.  
Detector noise will be taken to be homogeneous, with power 
spectra given by \cite{Knox95}
\begin{eqnarray}
\left.C_l^{\Theta\Theta}\right |_{\text{noise}}
&=&\left (\frac{\Delta_\Theta}{T_{\text{CMB}}}\right )^2
e^{l(l+1)\fwhm^2/8\ln 2},\nonumber \\
\left.C_l^{EE}\right |_{\text{noise}}
&=&\left.C_l^{BB}\right |_{\text{noise}} = 
\left (\frac{\Delta_P}{T_{\text{CMB}}}\right )^2
e^{l(l+1)\fwhm^2/8\ln 2},
\label{Eqn:HomogeneousNoise}
\end{eqnarray}
where $\Delta_\Theta$ and $\Delta_P$ characterize detector noise, and 
$\fwhm$ is the FWHM of the beam.  We employ the specifications
of a nearly ideal 
reference experiment, with 
$\Delta_\Theta = 1\mu\text{K-arcmin}$, $\Delta_P=\sqrt{2}\mu\text{K-arcmin}$,
and $\fwhm = 4'$ (see \cite{HuOkamoto02} for an exploration of noise properties). 
 
If we instead consider an ensemble of CMB fields 
lensed by a \emph{fixed} deflection field, the multipole covariance 
acquires off-diagonal terms 
and becomes
\begin{equation}
\avg{a_l^mb_{l'}^{m'}}{\Big|}_{\text{lens}}=
{C}_l^{ab} 
\delta_{ll'}\delta_{m-m'}(-1)^{m} + 
\sum_{LM}(-1)^M
\ThreeJ{l}{l'}{L}{m}{m'}{-M}f^\alpha_{lLl'}\phi_L^M,
\label{Eqn:LensAverage}
\end{equation}
where the subscript on the average indicates that we consider a fixed lensing 
field.  
$f^\alpha_{lLl'}$ are weights for the different quadratic pairs denoted by 
$\alpha$, given by
\begin{equation}
f^\alpha_{l_1Ll_2} = {}_{s_a}F_{l_1Ll_2}
\left [\even_{l_1l_2L}\tilde{C}_{l_2}^{ab}
+\odd_{l_1l_2L}\tilde{C}_{l_2}^{b\bar a} \right ]+{}_{s_b}F_{l_2Ll_1}
\left [\even_{l_1l_2L}\tilde{C}_{l_1}^{ab}
-\odd_{l_1l_2L}\tilde{C}_{l_1}^{a \bar b} \right ],
\label{Eqn:fDefinition}
\end{equation}
where $s_a$ and $s_b$ are the spins of the $a$ and $b$ fields respectively.
Specific forms for the six quadratic pairs are given in 
Table~\ref{Table:fForms}.
\begin{table}[htbp]
\begin{tabular}{c|c}
$\alpha$ & $f^\alpha_{l_1Ll_2}$\\
\hline
$\Theta\Theta$ & $\tilde{C}^{\Theta\Theta}_{l_1} {}_0 F_{l_2Ll_1}+
\tilde{C}^{\Theta\Theta}_{l_2} {}_0 F_{l_1Ll_2}$\\
$\Theta E$ & $\tilde{C}^{\Theta E}_{l_1}{}_2F_{l_2Ll_1}+
\tilde{C}^{\Theta E}_{l_2} {}_0 F_{l_1Ll_2}\text{, even}$\\
$EE$ & $\tilde{C}^{EE}_{l_1}{}_2 F_{l_2Ll_1}+
\tilde{C}^{EE}_{l_2}{}_2 F_{l_1Ll_2}\text{, even}$\\
$\Theta B$ & $i\tilde{C}^{\Theta E}_{l_1}{}_2 F_{l_2Ll_1}\text{, odd}$\\
$EB$ & $i\left [\tilde{C}^{EE}_{l_1}{}_2F_{l_2Ll_1}
-\tilde{C}^{BB}_{l_2}{}_2F_{l_1Ll_2}\right ]\text{, odd}$ \\
$BB$ & $\tilde{C}^{BB}_{l_1}{}_2F_{l_2Ll_1}+
\tilde{C}^{BB}_{l_2}{}_2F_{l_1Ll_2}\text{, even}$
\end{tabular}
\caption{Functional forms for $f^\alpha_{l_1Ll_2}$.  ``Even'' and ``odd'' 
indicate that the functions are non-zero only when $L+l_1+l_2$ is even or 
odd, respectively.}
\label{Table:fForms}
\end{table}

Because $\phi_L^M$ has a zero mean, the off-diagonal terms of the 
two-point correlations $\avg{a_l^mb_{l'}^{m'}}$ taken over a statistical 
ensemble would vanish.
However, in a given realization, we can construct an estimator for the 
deflections as a weighted sum over multipole pairs, and find weights that 
minimize the variance of the estimator.  We write a general weighted sum of 
multipole pairs as
\begin{equation}
d^\alpha{}_L^M=\frac{A^\alpha_L}{\sqrt{L(L+1)}}\sum_{l_1m_1}\sum_{l_2m_2}(-1)^M
\ThreeJ{l_1}{l_2}{L}{m_1}{m_2}{-M}g^\alpha_{l_1l_2}(L)
a_{l_1}^{m_1}b_{l_2}^{m_2},
\label{Eqn:EstimatorDefinition}
\end{equation}
where $a_l^m$ and $b_l^m$ are the observed CMB multipoles, $\alpha$ denotes 
the specific choice of $a$ and $b$, and the sum includes the diagonal 
($l_1=l_2$, $m_1=-m_2$) pieces.  

>From expression (\ref{Eqn:LensAverage}) for the average over a fixed lens 
realization, 
\begin{eqnarray}
\avg{d^\alpha{}_L^M}{\Big|}_{\text{lens}}&=&\frac{A^\alpha_L}{\sqrt{L(L+1)}}
\left [\sum_l\sqrt{\frac{2l+1}{2L+1}}g^\alpha_{ll}(L)
C_l^{ab} 
\delta_{L0}+\frac{\phi_L^M}{2L+1}\sum_{l_1l_2}g^\alpha_{l_1l_2}(L)
f^\alpha_{l_1Ll_2}
\right ].
\label{Eqn:EstimatorLensReduced}
\end{eqnarray}
where we have used the relations
\begin{eqnarray}
\sum_{m_1m_2}\ThreeJ{l_1}{l_2}{L}{m_1}{m_2}{M}
\ThreeJ{l_1}{l_2}{L'}{m_1}{m_2}{M'} &=& \frac{1}{2L+1}\delta_{LL'}\delta_{MM'},
\label{Eqn:ThreeJUnitarity}\\
\sum_{m}(-1)^{l+m}\ThreeJ{l}{l}{L}{m}{-m}{0} &=&
\sqrt{\frac{2l+1}{2L+1}}\delta_{L0}.
\label{Eqn:SumForDiagonal}
\end{eqnarray}
The diagonal terms in Eq. (\ref{Eqn:EstimatorDefinition}) only 
contribute to
the unobservable monopole piece and we hereafter implicitly consider
$L>0$ only.  
The normalization is set by the condition
\begin{equation}
\avg{d^\alpha{}_L^M} {\big|}_{\text{lens}} = \sqrt{L(L+1)}\phi_L^M
\label{Eqn:UnbiasedCondition}
\end{equation}
to be 
\begin{equation}
A^\alpha_L = L(L+1)(2L+1)\left \{
\sum_{l_1l_2}g^\alpha_{l_1l_2}(L)f^{\alpha}_{l_1Ll_2}
 \right \}^{-1}.
\label{Eqn:Normalization}
\end{equation}

We derive the minimum variance estimator by minimizing 
the Gaussian variance $\avg{d^\alpha{}_L^{M*}d^\alpha{}_L^M}$
with respect to $g^\alpha_{l_1l_2}(L)$ 
and find that
\begin{equation}
g^\alpha_{l_1l_2}(L)=\frac{C_{l_2}^{aa}C_{l_1}^{bb}f^{\alpha *}_{l_1Ll_2}
-(-1)^{L+l_1+l_2}C_{l_1}^{ab}C_{l_2}^{ab}f^{\alpha *}_{l_2Ll_1}}
{C_{l_1}^{aa}C_{l_2}^{aa}C_{l_1}^{bb}C_{l_2}^{bb}
-(C_{l_1}^{ab}C_{l_2}^{ab})^2}.
\label{Eqn:gFilter}
\end{equation}
Note that for $a=b$,
\begin{equation}
g^\alpha_{l_1l_2}(L)\rightarrow \frac{f^{\alpha *}_{l_1Ll_2}}
{2C_{l_1}^{aa}C_{l_2}^{aa}}, 
\label{Eqn:g(a=b)}
\end{equation}
and for $C_l^{ab}=0$ (e.g., for $\Theta B$ or $EB$),
\begin{equation}
g^\alpha_{l_1l_2}(L)\rightarrow\frac{f^{\alpha *}_{l_1Ll_2}}
{C_{l_1}^{aa}C_{l_2}^{bb}}.
\label{Eqn:g(Cl^ab=0)}
\end{equation}
The Gaussian noise covariance 
\begin{equation}
\avg{d^{\alpha }{}_L^{M*} d^\beta{}_{L'}^{M'}}\equiv 
\delta_{L,L'}\delta_{M,M'}\left [C_L^{dd} + N^{\alpha\beta}_L \right ]
\end{equation}
is given by
\begin{equation}
N^{\alpha\beta}_L=\frac{A_L^{\alpha*}A_L^\beta}{L(L+1)(2L+1)}
\sum_{l_1l_2}\left \{ g^{\alpha*}_{l_1l_2}(L)\left [C_{l_1}^{ac}C_{l_2}^{bd}
g^\beta_{l_1l_2}(L)+(-1)^{L+l_1+l_2}C_{l_1}^{ad}C_{l_2}^{bc}g^\beta_{l_2l_1}(L)
 \right ]\right \},
\label{Eqn:NoiseCovariance}
\end{equation}
with $\alpha=(ab)$, $\beta=(cd)$.  For $\alpha=\beta$, the above reduces 
simply to $N^{\alpha\alpha}_L = A^\alpha_L$.

Following the treatment of the flat sky case in \cite{HuOkamoto02}, we 
combine the measured quadratic estimators to further improve the signal to
noise by a forming minimum variance estimator
\begin{equation}
d^{\text{mv}}{}_L^M =\sum_\alpha w^\alpha(L)d^\alpha{}_L^M,
\label{Eqn:WeightedSum}
\end{equation}
with weights and variance given by
\begin{eqnarray}
w^\alpha(L)&=& 
{N^{\text{mv}}_L}
{\sum_\beta \left ({\bf N}^{-1}_L \right )^{\alpha\beta}},
\label{Eqn:MVWeighting}\\
N^{\text{mv}}_L& =& \frac{1}{\sum_{\alpha\beta}\left ( {\bf N}^{-1}_L
\right )^{\alpha\beta}}. \label{Eqn:MVNoise} \\
\end{eqnarray}
We will hereafter ignore contributions from the $BB$ estimator, since the primordial 
contributions to the $B$-mode power spectrum is expected to be small on scales
where the 
lensed multipoles are employed.
\begin{figure}[t!]
\myfigure{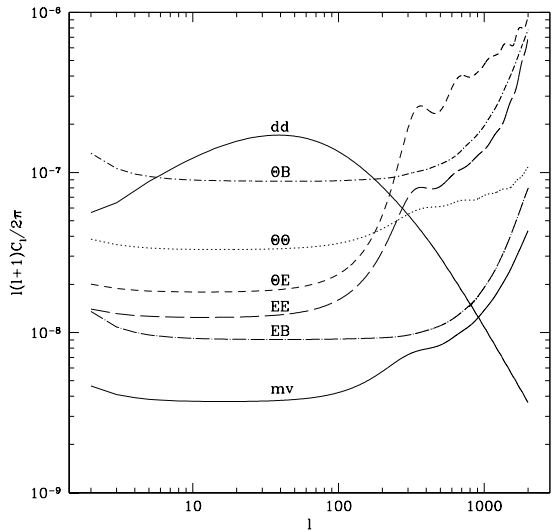}
\caption{Deflection and noise power spectra for the quadratic and minimum
variance estimators, assuming the noise properties of the reference 
experiment ($\Delta_\Theta=1\mu$K-arcmin; $\Delta_P = \sqrt{2}\mu$K-arcmin;
$\theta_{\rm FWHM}=4'$) and a fiducial $\Lambda$CDM cosmology with
with parameters
$\Omega_c = 0.3$, $\Omega_b=0.05$, $\Omega_\Lambda=0.65$, $h=0.65$, $n=1$,
$\delta_H=4.2\times 10^{-5}$ and no gravitational waves.}
\label{Fig:FullSkyNoise}
\end{figure}

We plot the noise power spectra for the five estimators, as well as the 
minimum variance estimator, in Fig.~\ref{Fig:FullSkyNoise}, assuming the 
noise properties of the reference experiment.

\section{Efficient Estimators}
\label{Sect:AngularSpaceEstimators}
The quadratic estimators involve both filtering and convolution 
in harmonic space.
It is useful in practice to express the convolution as a product
of the fields in angular space.   The estimators can then be constructed
using fast harmonic transform algorithms \cite{Healpix,MucNatVit97}. 
To simplify the
construction of the estimators we will assume $\tilde C_l^{BB} \ll \tilde C_l^{EE}$    
as is appropriate for the standard cosmology.
Aside from the $EB$ estimator, derived in \cite{HuOkamoto02} under the
flat sky approximation, the angular space estimators
involving polarization are new to this work.

Generalizing the construction in \cite{Hu01a} for the $\Theta\Theta$ 
estimator, consider the fact that lensing correlates
the (lensed) temperature and polarization fields to the their (unlensed) angular gradients.   
We show in Appendix~\ref{Appendix:SpinSFunctions} that the all-sky
analog to the gradient operation on a spin-$s$ field is  $\partial_i \rightarrow D_i$.
The quadratic estimator is then built out of the general operation on
two fields $X(\vh{n})$ and $Y(\vh{n})$
\begin{equation}
P[X(\vh{n}),Y(\vh{n})] \equiv - D^i [ X(\vh{n}) D_i Y(\vh{n}) ]\,.
\label{Eqn:gradientcorr}
\end{equation}

The properly normalized estimators then take the form
\begin{equation}
\hat d^\alpha{}_L^M=\frac{\hat A^\alpha_L}{\sqrt{L(L+1)}}\int d\vh{n}  Y_L^{M*} (\vh{n})  e^\alpha(\vh{n})\,,
\label{Eqn:angularestimator}
\end{equation}
where 
\begin{align}
e^{\Theta\Theta}(\vh{n}) &= P[{}_0 A_\Theta, {}_0 A_{\Theta\Theta}]\,,\nonumber\\
e^{\Theta E}    (\vh{n}) &= \frac{1}{2} \left( 
P[{}_{+2} A_E, {}_{-2} A_{\Theta E}] + P[{\rm cc}] \right) + P[{}_0 A_\Theta, {}_0 A_{E\Theta}]\,, \nonumber\\
e^{\Theta B}    (\vh{n}) &= \frac{1}{2} \left( P[ {}_{+2} A_{iB}, {}_{-2} A_{\Theta E}] + P[{\rm cc}] \right) \,,\nonumber\\
e^{E E}    (\vh{n}) &= \frac{1}{2} \left( P[ {}_{+2} A_E, {}_{-2} A_{EE} ] + P[{\rm cc}] \right) \,,\nonumber\\
e^{E B}    (\vh{n}) &= \frac{1}{2} \left( P[ {}_{+2} A_{iB},{}_{-2} A_{EE}] + P[{\rm cc}] \right) \,,
\end{align} 
where cc denotes the operation with the complex conjugates of the fields and
the filtered fields themselves are given by the general prescription 
\begin{align}
{}_{\pm s} A_{X}(\vh{n}) &= \sum_{l m} \frac{1}{C_l^{XX}} X_l^m {}_{\pm s} Y_l^m(\vh{n}), \nonumber\\
{}_{\pm s} A_{XY}(\vh{n}) &= \sum_{l m} \frac{\tilde C_l^{XY}}{C_l^{XX}} X_l^m {}_{\pm s} Y_l^m(\vh{n}) .
\end{align}
We omit the $\alpha=BB$ estimator under the assumption that the unlensed $B$-power
is small at high multipoles.  

\begin{figure}[tbhp]
\myfigure{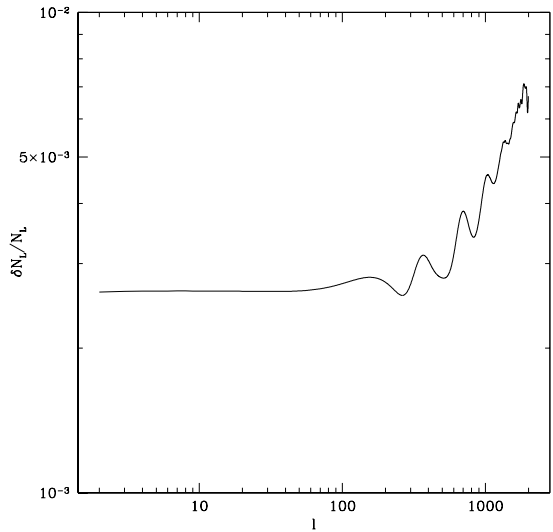}
\caption{Fractional difference between the approximate and the minimum variance
$\Theta E$ estimators, defined as $\delta N_L/N_L\equiv 
(N_L^{\text{approx.}}-N_L^{\text{mv}})/N_L^{\text{mv}}$.}
\label{Fig:TENoise}
\end{figure}

It is straightforward to verify that all of the estimators are the same as the
harmonic space ones $\hat d^\alpha{}_L^M = d^\alpha{}_L^M$ with $\hat A_L^\alpha = A_L^\alpha$ 
except for $\Theta E$.  Here the weights on the
multipole combination are 
\begin{align}
\hat g_{l_1 l_2}^{\Theta E} = 
\frac{f^{\Theta E}_{l_1 L l_2}}{C_{l_1}^{\Theta\Theta} C_{l_2}^{EE}}
\end{align}
and are slightly non-optimal compared with the minimum variance weighting.  
Furthermore $\hat N_L \ne \hat A_L$ and they must be calculated separately.
However a direct calculation of the noise spectrum
through Eqn.~(\ref{Eqn:NoiseCovariance})
shows that the differences are less than $1\%$, and 
essentially indistinguishable from the minimum variance estimator 
(see Fig.~\ref{Fig:TENoise}).  

These estimators may therefore be used in place of a direct multipole summation
for efficient lens reconstruction.
The gradient operations in Eqn.~(\ref{Eqn:gradientcorr}) are efficiently evaluated
in harmonic space since their action on spin harmonics simply raises and lowers
the spin index in accordance with Eqn.~(\ref{Eqn:spingradient}).

\section{Discussion}
\label{Sect:Conclusion}

Counterintuitively, the gravitational lensing of the CMB temperature and polarization
fields is a small scale manifestation of the very large scale properties of the
intervening mass distribution.  It therefore requires very challenging,
high angular resolution ($< 10'$) but wide-field surveys ($>$ few degrees) to exploit.  
We have provided expressions for quadratic estimators of the lensing potential
valid on the entire sky, as well as the expected noise covariances for the
estimators.  
As expected, on small angular scales ($L\ge 100$), the 
flat sky approximations differ from the full sky expressions
by less than $\sim 1\%$, indicating that the flat sky approximations is adequate.
This regime is however not where the signal-to-noise peaks. 

We have also provided a practical means of implementing these estimators
using fast harmonic transforms, either with spherical harmonics or Fourier harmonics, 
to perform the required harmonic convolutions and filtering.
We have shown that even the approximate $\Theta E$ estimator
has a noise performance that is essentially indistinguishable from the
minimum variance estimator. 
These techniques should provide a means to study the impact of real world issues 
such as finite-field, inhomogeneous noise, and foregrounds on the science of
CMB lensing.

\begin{acknowledgments}
This work was supported by
NASA NAG5-10840 and the DOE OJI program.
\end{acknowledgments}

\appendix
\section{Tensor Representation of Spin Weighted Functions}
\label{Appendix:SpinSFunctions}

We clarify the relation between spin-$s$ functions and tensor quantities on
the sphere, and derive the relation between spin raising and lowering operators
and covariant derivatives on the sphere.

Suppose we construct an orthonormal basis $(\vh{e}_1(\vh{n}),
\vh{e}_2(\vh{n}))$ at each point on the sphere, with $\vh{n}$ denoting the 
outward-facing normal vector.  We define a local rotation as
a right-handed rotation of the basis vectors $(\vh{e}_1,\vh{e}_2)$ by an angle 
$\psi$ around the vector $\vh{n}$, so that the new basis vectors 
$(\vh{e}_{1'},\vh{e}_{2'})$ are related to the original vectors by the 
transformation
\begin{eqnarray}
\vh{e}_{1'}&=&\cos\psi \vh{e}_1 + \sin\psi \vh{e}_2,\nonumber \\
\vh{e}_{2'}&=&-\sin\psi \vh{e}_1 + \cos\psi \vh{e}_2.
\label{Eqn:BasisRotation}
\end{eqnarray}
A function ${}_sf(\vh{n})$ is said to carry a spin-weight $s$ if, under the
rotation (\ref{Eqn:BasisRotation}), the function transforms as 
${}_sf(\vh{n})\rightarrow e^{-is\psi}{}_sf(\vh{n})$.  This convention conforms
to \cite{ZaldarriagaSeljak97}, and defines rotations in a sense opposite to 
that in \cite{Goldberg67,NewmanPenrose66}.  

We define vectors $\vm$ and $\vp$ with respect to the 
basis $(\vh{e}_1,\vh{e}_2)$ according to
\begin{eqnarray}
\vm &=& \frac{1}{\sqrt{2}}\left [\vh{e}_1 + i \vh{e}_2\right ],\\
\vp &=& \frac{1}{\sqrt{2}}\left [\vh{e}_1 - i \vh{e}_2\right ],
\label{Eqn:ProjectionVectors}
\end{eqnarray}
which have the property that 
\begin{eqnarray}
\vm\cdot\vm &=& \vp\cdot\vp = 0 , \nonumber \\
\vm\cdot\vp &=& 1.
\label{Eqn:Orthonormality}
\end{eqnarray}
Given a vector field $\vc{v}(\vh{n})$, it can easily be shown that the 
quantities $\vc{v}\cdot\vm$ and $\vc{v}\cdot\vp$ transform as spin $1$ and
$-1$ objects, respectively, so that $\vm$ and $\vp$ act as spin projection
vectors.  More generally, given a rank-$s$ tensor $T_{i_1\ldots i_s}$,
the quantity $T_{i_1\ldots i_s}\vm^{i_1}\cdots\vm^{i_s}$ transforms
as a spin-$s$ object, since under the rotation (\ref{Eqn:BasisRotation}), each 
factor of $\vm^{i_n}$ contributes a phase $e^{-is\psi}$.

The spin-$s$ functions ${}_sf(\vh{n})$ therefore also provide 
a complete basis for the totally symmetric trace-free portion of a rank-$s$ tensor
\begin{equation}
T_{i_1\ldots i_s}={}_sf(\vh{n})\vp_{i_1}\cdots\vp_{i_s} 
+{}_{-s}f(\vh{n})\vm_{i_1}\cdots\vm_{i_s},
\label{Eqn:RankSBoth}
\end{equation}
where the trace-free condition refers to the vanishing under contraction of
any two indices in the tensor.  
For example, the polarization tensor can be written as 
\begin{equation}
{\mathcal P}_{ij} = {}_{+2} A(\vh{n})\vp_i\vp_j + {}_{-2} A(\vh{n})\vm_i\vm_j .
\label{Eqn:PolarizationTrep}
\end{equation}

Covariant differentiation of such a tensor is related to the raising and lowering
of the spin weight \cite{Goldberg67}:  
\begin{eqnarray}
\nabla_k T_{i_1\ldots i_s}&=&\left [\partial_k{}_sf(\vh{n})\right ]
\vp_{i_1}\cdots\vp_{i_s} + {}_sf(\vh{n})\nabla_{(k}\vp_{i_1}\cdots\vp_{i_s)}
\nonumber \\
&+&\left [\partial_k{}_{-s}f(\vh{n})\right ]
\vm_{i_1}\cdots\vm_{i_s} + {}_{-s}f(\vh{n})\nabla_{(k}\vm_{i_1}\cdots
\vm_{i_s)}.
\label{Eqn:CovariantDerivative}
\end{eqnarray}
We evaluate the covariant derivatives $\nabla_i\vp_j$ etc. explicitly in the
spherical basis with coordinates $(\theta,\varphi)$, yielding
\begin{eqnarray}
\nabla_\theta\vp_\theta &=& \nabla_\theta\vp_\varphi = 0, \nonumber\\
\nabla_\varphi\vp_\theta &=& \frac{i}{\sqrt{2}}\cos\theta, \nonumber\\
\nabla_\varphi\vp_\varphi &=& \frac{1}{\sqrt{2}}\sin\theta\cos\theta,
\label{Eqn:ExplcitCovDerivs}
\end{eqnarray}
with those for $\vm$ given as complex conjugates of the above.
Using these, it can be shown that 
\begin{eqnarray}
\vp^j\nabla_i\vp_j &=& \vm^j\nabla_i\vm_j = 0,\nonumber\\
\vm^j\nabla_i\vp_j &=& -\vp^j\nabla_i\vm_j = J_i, 
\label{Eqn:CovInJ}
\end{eqnarray}
where 
\begin{equation}
J_\theta = 0, \qquad J_\varphi = i\cos\theta,
\label{Eqn:JEvaluation}
\end{equation}
in spherical coordinates.  The covariant derivative of $T_{i_1\ldots i_s}$
is therefore given by
\begin{equation}
\nabla_kT_{i_1\ldots i_s} = \left [ D_k\,
{}_sf(\vh{n})\right]\vp_{i_1}\cdots\vp_{i_s}+
\left[  D_k \, {}_{-s}f(\vh{n})\right]
\vm_{i_1}\cdots\vm_{i_s},
\label{Eqn:CovDerivFinal}
\end{equation}
where 
we define the spin-dependent gradient operator as 
\begin{eqnarray}
D_i&\equiv& \partial_i + sJ_i .
\label{Eqn:GradientOp}
\end{eqnarray}
A covariant derivative $\nabla_i$ operating on the spin-$s$ piece of a tensor
is equivalent to a gradient operation $D_i$ on its spin-$s$ weighted 
representation.
As an example, the components of the covariant derivative of the
polarization tensor 
\begin{eqnarray}
\vm^i\vm^j\nabla_k {\mathcal P}_{ij}&=&D_k[{}_2 A(\vh{n})],\nonumber \\
\vp^i\vp^j\nabla_k {\mathcal P}_{ij}&=&D_k[{}_{-2} A(\vh{n})].
\label{Eqn:DkExample}
\end{eqnarray}

The gradient operator $D_k$ is related to spin raising and lowering operators.
Using the expressions (\ref{Eqn:JEvaluation}) and expressing the operator $D_k$
in the $(\vm,\vp)$ basis, we obtain the desired relations
\begin{eqnarray}
D_i\, [{}_sf(\vh{n})]
&=&-\frac{1}{\sqrt{2}}\left \{\left [\edth {}_sf(\vh{n})\right ]\vp_i 
+ \left [\baredth {}_sf(\vh{n})\right ]\vm_i \right \}.
\label{Eqn:DerivToSpinLadder}
\end{eqnarray}
By virtue of the rotational properties of ($\vm$, $\vp$), the ladder
operators $\edth$ and $\baredth$, defined by 
\cite{Goldberg67,NewmanPenrose66}
\begin{eqnarray}
\edth{}_sf(\theta,\varphi) &=& -\sin^s\theta\left [ 
\frac{\partial}{\partial\theta}+i\csc\theta\frac{\partial}{\partial\varphi}
\right ]\sin^{-s}\theta{}_sf(\theta,\varphi), \label{Eqn:Lowering}\\
\baredth{}_sf(\theta,\varphi)&=&-\sin^{-s}\theta\left [ 
\frac{\partial}{\partial\theta}-i\csc\theta\frac{\partial}{\partial\varphi}
\right ]\sin^s\theta{}_sf(\theta,\varphi),\label{Eqn:Raising}
\end{eqnarray}
raise and lower the spin weight by $1$.  For example, the gradient operation
on the spin-$s$ spherical harmonic yields
\begin{equation}
D_i [{}_s Y_l^m ]
= -\frac{1}{\sqrt{2}} \left( [(l-s)(l+s+1)]^{1/2} {}_{s+1}Y_l^m  \vp_i
	- [(l+s)(l-s+1)]^{1/2} {}_{s-1}Y_l^m \vm_i\right).
\label{Eqn:spingradient}
\end{equation}
Note that the inner product of two gradients
\begin{eqnarray}
[D^i\, {}_{s_1} f_1(\vh{n})]
[D_i\, {}_{s_2} f_2(\vh{n})]
= 
\frac{1}{2}\left \{ 
[\baredth\, {}_{s_1} f_1(\vh{n})]\, [\edth{}_{s_2}f_2 (\vh{n}) ]
+[\edth {}_{s_1} f_1(\vh{n})]\, [\baredth \, {}_{s_2}f_2(\vh{n}) ]\right \}
\label{Eqn:gradientdot}
\end{eqnarray}
leaves the total spin-weight of the product unchanged.

Inverting the relation (\ref{Eqn:DerivToSpinLadder}), we obtain the ladder operators in the
tensor representation
\begin{eqnarray}
\edth {}_sf(\vh{n})&=&-\sqrt{2}\vm^j\vm^{i_1}\cdots\vm^{i_s}
\nabla_jT_{i_1\ldots i_s},\label{Eqn:EdthInTensor}\\
\baredth {}_sf(\vh{n})&=&-\sqrt{2}\vp^j\vm^{i_1}\cdots\vm^{i_s}
\nabla_jT_{i_1\ldots i_s},\label{Eqn:BaredthInTensor}
\end{eqnarray}
for $s\ge 0$, and with $\vm^{i_n}$ replaced by $\vp^{i_n}$ for $s<0$.
This relationship was first proven in \cite{Goldberg67}, albeit with 
a different sign convention.

\begin{figure}[b!]
\myfigure{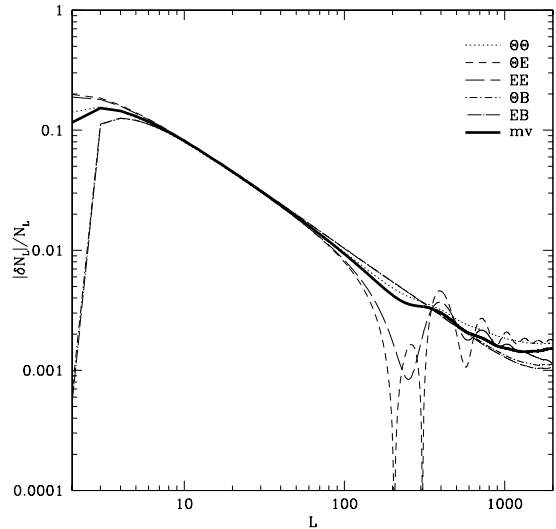}
\caption{Fractional differences $\delta N_L/N_L$ between the noise in the 
flat sky and full sky estimators, calculated for the reference experiment.}
\label{Fig:Comparison}
\end{figure}

\section{Flat-sky Approximation}
\label{Appendix:FlatSky}

The all-sky estimators derived in Sect. \ref{Sect:QuadraticEstimators}
reduce to the flat sky estimators, based on
Fourier harmonics of the fields in the small angle limit.  
Here we explicitly show this correspondence.

The full sky harmonics in multipole space $(l,m)$
are related to the flat sky harmonics in Fourier space 
$\vc{l} = l \cos \varphi_l \vh{x} +
l \sin \varphi_l \vh{y}$ by \cite{Whiteetal99,Hu00}
\begin{equation}
X(\vc{l})=\sqrt{\frac{4\pi}{2l+1}}\sum_mi^{-m}X_l^me^{im\varphi_l}.
\label{Eqn:FlattoFull}
\end{equation}
Rewriting Eq. (\ref{Eqn:EstimatorDefinition}) using the above, 
\begin{eqnarray}
d^\alpha(\vc{L})&\approx&\frac{A_L^\alpha}{\sqrt{L(L+1)}}\sum_M(-i)^M
e^{iM\varphi_L}\sum_{l_1m_1}\sum_{l_2m_2}i^{m_1+m_2}g^\alpha_{l_1l_2}(L)
\ThreeJ{l_1}{l_2}{L}{m_1}{m_2}{-M}
\nonumber\\
&\times&
\sqrt{\frac{(2l_1+1)(2l_2+1)}{4\pi (2L+1)}}
\int\frac{d\varphi_1}{2\pi}\frac{d\varphi_2}{2\pi}
e^{-im_1\varphi_1}e^{-im_2\varphi_2}
a(\vc{l}_1)b(\vc{l}_2).
\label{Eqn:FlatEstimator1}
\end{eqnarray}
To go further, we can utilize the approximation \cite{Hu00}
\begin{equation}
\ThreeJ{l_1}{L}{l_2}{2}{0}{-2}\approx\ThreeJ{l_1}{L}{l_2}{0}{0}{0}\times
\left\{\begin{array}{cc}
\cos 2(\varphi_{l_2}-\varphi_{l_1}), & L+l_1+l_2=\text{even, }\\
i\sin 2(\varphi_{l_2}-\varphi_{l_1}), & L+l_1+l_2=\text{odd,}
\end{array}\right .
\label{Eqn:ThreeJApprox}
\end{equation}
with the trigonometric functions defined through the cosine and sine rules,
and the 3-j symbol on the rhs for the odd case represents a continuation 
of the analytic expression for the even case
\begin{equation}
\ThreeJ{l_a}{l_b}{l_c}{0}{0}{0} = (-1)^{l/2}\frac{\left (\frac{l}{2}\right )!}
{\left (\frac{l}{2}-l_a \right )!\left (\frac{l}{2}-l_b \right )!
\left (\frac{l}{2}-l_c \right )!}
\left [\frac{(l-2l_a)!(l-2l_b)!(l-2l_c)!}{(l+1)!} \right ]^{1/2},
\label{Eqn:ThreeJExpression}
\end{equation}
where $l=l_a+l_b+l_c$.

Furthermore, in the limit
$l_1,l_2,L\gg 1$, 
\begin{equation}
\frac{1}{2}\left [L(L+1)+l_2(l_2+1)-l_1(l_1+1)\right ] \approx 
\vc{L}\cdot\vc{l}_2,
\end{equation}
and so we may absorb the geometric factors in
${}_sF_{l_1Ll_2}$ as 
\begin{equation}
{}_sF_{l_1Ll_2}\approx \sqrt{\frac{(2L+1)(2l_1+1)(2l_2+1)}{4\pi}}
\ThreeJ{l_1}{l_2}{L}{0}{0}{0}{}_s \bar F_{l_1Ll_2},
\label{Eqn:FinMu}
\end{equation}
with
\begin{eqnarray}
{}_s \bar F_{l_1Ll_2}&\equiv& 
\vc{L}\cdot\vc{l}_2
\times\left \{\begin{array}{cc}
1, & s=0,\\
\cos 2(\varphi_{l_2}-\varphi_{l_1}), & s=2, L+l_1+l_2=\text{even, }\\
i\sin 2(\varphi_{l_2}-\varphi_{l_1}), & s=2, L+l_1+l_2=\text{odd.}
\end{array}\right . \nonumber \\
\label{Eqn:MuDefn}
\end{eqnarray}

The approximations for $f^\alpha_{l_1Ll_2}$ and $g^\alpha_{l_1l_2}(L)$ can 
likewise be written as
\begin{eqnarray}
f^\alpha_{l_1Ll_2}&\approx&\sqrt{\frac{(2L+1)(2l_1+1)(2l_2+1)}{4\pi}}
\ThreeJ{l_1}{l_2}{L}{0}{0}{0}\bar{f}^\alpha_{l_1Ll_2},\nonumber \\
g^\alpha_{l_1l_2}(L)&\approx&\sqrt{\frac{(2L+1)(2l_1+1)(2l_2+1)}{4\pi}}
\ThreeJ{l_1}{l_2}{L}{0}{0}{0}\bar{g}^\alpha_{l_1l_2}(L),
\label{Eqn:BarQuantities}
\end{eqnarray}
where $\bar{f}^\alpha_{l_1Ll_2}$ and $\bar{g}^\alpha_{l_1l_2}(L)$ are defined
as the unbarred quantities with ${}_sF_{l_1Ll_2}$ replaced by 
${}_s \bar F_{l_1Ll_2}$.  
We will also utilize the relation between plane waves and spherical harmonics,
\begin{equation}
e^{i\vc{l}\cdot\vh{n}}\approx \sqrt{\frac{2\pi}{l}}\sum_m i^mY_l^m
e^{-im\varphi_l}.
\label{Eqn:PlaneWave}
\end{equation}

Using Eq. (\ref{Eqn:BarQuantities}) to rewrite Eq. (\ref{Eqn:FlatEstimator1}),
and applying relations (\ref{Eqn:ThreeJtosYlm}) and (\ref{Eqn:PlaneWave}), the 
estimator becomes
\begin{eqnarray}
d^\alpha(\vc{L})&\approx&\frac{A_L^\alpha}{\sqrt{L(L+1)(2L+1)}}
\sum_{l_1 l_2} \bar{g}^\alpha_{l_1l_2}(L)
\sqrt{\frac{(2l_1+1)(2l_2+1)}{4\pi}}\nonumber \\
&\times & \int\frac{d\varphi_1}{2\pi}\frac{d\varphi_2}{2\pi}
\sqrt{\frac{l_1l_2 L}{(2\pi)^3}}
a(\vc{l}_1)b(\vc{l}_2) 
\int d\vh{n}
e^{i(\vc{l}_1+\vc{l}_2+\vc{L})\cdot\vh{n}}.
\label{Eqn:FlatEstimator2}
\end{eqnarray}
Taking $l_1,l_2,L\gg 1$, the above reduces to
\begin{eqnarray}
d^\alpha(\vc{L})&\approx&\frac{A_L^\alpha}{L}\int\frac{d^2\vc{l}_1}{(2\pi)^2}
\bar{g}^\alpha_{l_1l_2}(L)a(\vc{l}_1)b(\vc{l}_2), \quad \vc{l}_2 = \vc{L}-\vc{l}_1
\label{Eqn:FlatEstimatorFinal}
\end{eqnarray}
with $\bar{g}^\alpha_{l_1l_2}(L)$ corresponding to the filters
$F_\alpha(\vc{l}_1,\vc{l}_2)$ in \cite{HuOkamoto02}.  
The normalization $A^\alpha_L$ reduces to the flat sky 
expression in \cite{HuOkamoto02} in a similar fashion, by using the 
approximations (\ref{Eqn:BarQuantities}) to relate the full sky quantities
to trigonometric functions on the flat sky. 

It is simple to show that the efficient all-sky estimator in 
Eqn.~(\ref{Eqn:angularestimator}) reduces to efficient flat sky
estimators with the replacements $D_i \rightarrow \partial_i$ in Eqn.~(\ref{Eqn:gradientcorr})
and the spherical harmonic transform in 
Eqn.~(\ref{Eqn:angularestimator}) with a Fourier transform.  Under the 
assumption that
$\tilde C_l^{BB} \ll \tilde C_l^{EE}$, they again reproduce the properties of the
minimum variance quadratic estimators in Eqn.~(\ref{Eqn:FlatEstimatorFinal}) and allow
fast Fourier transform techniques to be employed in their construction.

Fig.~\ref{Fig:Comparison} shows 
fractional differences between the noise in flat sky estimators derived in 
\cite{HuOkamoto02} and the noise in full sky estimators, defined as
$\delta N_L^\alpha/N_L^\alpha\equiv
(N_L^{\alpha(\text{flat})}-N_L^{\alpha(\text{full})})/
N_L^{\alpha (\text{full})}$.  Because most of the information comes from
multipole pairs at high multipole moments, the flat sky expressions deviate
at less than $\sim 1\%$ for $L> 200$, mainly in the direction of overestimating the 
noise.

\end{document}